\def\prd{Phys. Rev. D}

\def\apj{Astrophys. J.}
\def\apjl{Astrophys. J. Lett.}
\def\apjs{Astrophys. J.Suppl.}
\def\mnras{Mon. Not. R. Astr. Soc.}
\def\aap{Astr. Astrophys.}

\def \<{\langle}
\def \>{\rangle}

\newcommand{\ra}{\;\raise1.0pt\hbox{$'$}\hskip-6pt\partial\;}
\newcommand{\lo}{\;\overline{\raise1.0pt\hbox{$'$}\hskip-6pt\partial}\;}

\newcommand{\degree}{^\circ}

\newcommand{\exptt}[1]{\langle#1\rangle}

\newcommand{\Abs}{\abstract}
\newcommand{\Ack}{\acknowledgments}
\newcommand{\mktt}{\maketitle}

\documentclass[a4paper,11pt]{article}
\pdfoutput=1

\usepackage{jcappub,graphicx,epsfig,natbib,color,times,bm,amsmath,multirow,hyperref,amssymb,mathtools,mathrsfs}
\usepackage[ddmmyyyy,hhmmss]{datetime}

\begin{document}

\title{Pixel domain multi-resolution minimum variance painting of CMB maps}

\author[a]{Hao Liu}\emailAdd{ustc\_liuhao@163.com}

\affiliation[a]{School of Physics and Material Science, Anhui University, 111
Jiulong Road, Hefei, Anhui, China 230601.}

\Abs{

This work introduces an unbiased minimum variance painting of the pixel domain
CMB maps for both the missing and available sky regions. In the missing
region, it generates CMB realizations that have identical statistical
properties to the expected CMB signal; and in the available region, it
significantly alleviate the unwanted residuals, e.g., the residual of
EB-leakage. The time cost of this method follows $\sim$$ O(N_{side}^3)$
($N_{side}$ is the map resolution parameter), which is similar to the fast
spherical harmonic transforms and is much better than the traditional minimum
variance method that follows $\sim$$O(N_{side}^6)$.  This method is the basis
of a series of minimum variance estimations for analyzing CMB datasets, and
more works based on it will follow in the future.

}

\mktt

\section{Introduction}
\label{sec:intro}

In the measurement of cosmic microwave background (CMB) anisotropy, a few key
problems were constantly studied from the beginning until the present day. For
example, how to compute the full sky CMB angular power spectrum (APS) from
various cosmological models and vice versa; and how to estimate the true CMB
signal and APS from incomplete and contaminated sky maps. In this work, we
focus on the latter, especially the estimation and reconstruction of pixel
domain CMB signals, because it usually makes following CMB analysis much
easier.

For the case of CMB temperature, the best unbiased solution of this kind of
problem was given by~\citep{1997PhRvD..55.5895T}, which is usually referred to
as the fisher estimator, a method that finds the CMB APS with the maximum
likelihood condition. It was also mentioned in~\citep{1997PhRvD..55.5895T},
that the maximum likelihood and minimum variance conditions are equivalent for
the Gaussian isotropic CMB signal; thus, one is free to start from either
side.

The main problem of the fisher estimator is the computational cost. According
to the tests in~\cite{2014MNRAS.440..957M}, a full implementation of the
fisher estimator at a relatively low resolution ($N_{side}=64$) of the HEALPix
pixelization scheme~\cite{2005ApJ...622..759G}  already requires about
$4\times10^5$ CPU hours and 19 TB memory, which is a heavy task even for a
super cluster. For higher resolutions, the time cost scales as
$O(N_{side}^6)$; thus, the computation is essentially impossible. Therefore,
many alternative methods were developed. Some of them keep pursuing the
maximum likelihood (or minimum variance), but manage to do the job in a faster
way (\cite{2004ApJS..155..227E, 2004ApJ...609....1J, 2004PhRvD..70h3511W,
2005PhRvD..71j3002C, 2008ApJ...676...10E, 2009MNRAS.400..463G,
2009ApJ...697..258J}, and a review can be found in section~5
of~\cite{2019arXiv190909375G}); and some others choose to accept sub-optimal
results, but as an exchange, the time cost can be significantly reduced, e.g.,
the pseudo-$C_\ell$ method~\cite{2002ApJ...567....2H, 2001PhRvD..64h3003W} and
the Spice estimator~\cite{2004MNRAS.350..914C}. Overall speaking, the former
still requires to use a super cluster, especially for the case of higher
resolutions or higher-$\ell$s; and the latter is the currently most practical
way to estimate the high-$\ell$ CMB APS, but the quality of the result is
relatively lower, and especially, it cannot deal with a pixel-domain
estimation.

For the case of polarization, a key problem that hinders the minimum variance
solution is the leakage from the E-mode polarizations to the B-mode
polarizations, called the EB-leakage~\cite{2002PhRvD..65b3505L, Bunn:2002df}.
However, in our recent work~\cite{Liu_2019_EB_general}, it was shown how to
explicitly correct the EB-leakage in the pixel domain, and with some clearly
defined preconditions, the results are proven to be the best ones. Thus, it is
about time to design a series of minimum variance solution for both the
temperature and polarization, and from map to the final cosmological
parameters. Given the solution of EB-leakage, the most urgent problem at the
moment is to greatly reduce the computational cost of the pixel domain minimum
variance estimation, which is the basis of all following works in this series,
and will be the main goal of this work.

Meanwhile, by greatly reducing the time cost of pixel domain minimum variance
estimation, several interesting possibilities are also turned on, especially
for studying the large-scale and low-$\ell$ problems. For example, it will
become possible to change the sky mask pixel-by-pixel to investigate the
impact upon all kinds of large scale anomalies, in order to find out a
possible association to a specific sky region.

The outline of this work is as follows: In section~\ref{sec:math basis}, we
explain some mathematical basis of the method, and the technical details of
implementation are introduced in section~\ref{sec:implementation}. The
implementation and performance are tested and illustrated in
section~\ref{sec:results}, and a conclusion and discussion is given in
section~\ref{sec:conclusion}.

\section{The mathematical basis}
\label{sec:math basis}

The multi-resolution minimum variance painting (MrMVP) is based on the ideal
of constrained in-painting by~\cite{2012ApJ...750L...9K}, but some similar
ideas can also be found in~\cite{2012arXiv1211.0585E, 2013A&A...549A.111E,
2017MNRAS.468.1782K}. Basically, assume we have an available dataset $\bm{X} =
\bm{S}+\bm{N}$, where $\bm{S}$ is the true signal and $\bm{N}$ is the noise.
For convenience, here $\bm{N}$ includes all unwanted components like the
noise, systematics and various residuals. The main goal of painting is to
estimate $\bm{S}$ from $\bm{X}$ in both the missing and available regions,
including the expectation and constrained realizations. Customarily, when we
focus on the reconstruction of the missing region, the operation is called an
in-painting.

A basic in-painting method only focus on estimating the expectation in the
missing region, like the diffusive in-paint method~\citep{2014A&A...571A..24P,
2016A&A...594A..17P}; whereas a complete in-painting method will consider the
expectation and constrained realizations at the same time, like the one
in~\cite{2012ApJ...750L...9K}. With the well known theory of multi-variant
minimum variance regression (for example, see appendix A
of~\cite{Liu_2019_EB_general}), if we only want to solve for the expectation,
then the result is given by the following unbiased minimum variance solution:
\begin{eqnarray}\label{equ:min-var basic}
\bm{\widetilde{S}} &=& \bm{P Q}^{-1} \bm{X},
\end{eqnarray}
where $\bm{\widetilde{S}}$ is the pixel domain expectation of $\bm{S}$ and
$\bm{P} = \<\bm{XS}^T\>$, $\bm{Q} = \<\bm{XX}^T\> $ are the covariance
matrices of $\bm{X}$-with-$\bm{S}$ and $\bm{X}$-with-$\bm{X}$, respectively.
If we further assume $\bm{S}$ and $\bm{N}$ are uncorrelated (which is usually
true), then the above equation becomes
\begin{eqnarray}
\bm{P} = \exptt{\bm{SS}^T},~~\bm{Q} = \exptt{\bm{SS}^T} + \<\bm{NN}^T\>,
\end{eqnarray}
and eq.~(\ref{equ:min-var basic}) becomes the well known Wiener filter. Note
that in eq.~(\ref{equ:min-var basic}), if we are dealing with the missing
region, then $\bm{X}$ and $\bm{S}$ belong to different regions, but $\bm{P}$
is not zero, which makes the estimation of the missing region possible.

It is important to emphasis that $\bm{\widetilde{S}}$ is an ``expectation''
but not a ``realization''; such that the true signal $\bm{S}$ is always
different to $\bm{\widetilde{S}}$. In fact, in the view of statistics, the
true signal $\bm{S}$ is included as a member of the ensemble $\{\bm{S}_i\}$
containing all possible realizations, and $\bm{S}$ is not superior compared to
any other members. Therefore, the next immediate question is how to generate
the constrained ensemble $\{\bm{S}_i\}$, which is important because
$\bm{S}_i$, as fullsky realizations of the CMB signal, make many estimations
extremely convenient, as also mentioned in~\cite{2012ApJ...750L...9K}.

The generation of ensemble $\{\bm{S}_i\}$ is based on the following constraint
and prior conditions:
\begin{enumerate}
\item \label{itm:1} Constraint: $\bm{S}_i$ has to be consistent with the
actual dataset $\bm{X}$.
\item \label{itm:2} Prior condition: In case of CMB, $\bm{S}_i$ should be
Gaussian and isotropic.
\item \label{itm:3} Prior condition: $\bm{P}$ and $\bm{Q}$ are determined
either by the prior APS or by an iterative solution.
\end{enumerate}
With the above constraint and prior conditions, the ensemble $\{\bm{S}_i\}$
can be generated from the key equation provided in~\cite{1991ApJ...380L...5H,
2012ApJ...750L...9K}. For convenience of reading, it is retyped below using
the above notations:
\begin{eqnarray}\label{equ:CR}
\bm{S}_i = \bm{\widetilde{S}} + \bm{s}_i-\bm{P Q}^{-1}\bm{x}_i,
\end{eqnarray}
where $\bm{S}_i$ is a constrained realization; $\bm{s}_i$ is the $i$-th
unconstrained fullsky realization generated with the prior conditions
\ref{itm:2}--\ref{itm:3}, $\bm{x}_i$ is $\bm{s}_i$ plus noise, and
$\bm{\widetilde{S}}$ is the expectation given by~\ref{equ:min-var basic} that
carries the constraint~\ref{itm:1}. The physical meaning of the above equation
is: a properly constrained realization is the expectation $\bm{\widetilde{S}}$
solved from the real data, plus the difference between an unconstrained
realization $\bm{s}_i$ and the expectation $\bm{P Q}^{-1}\bm{x}_i$ solved from
$\bm{s}_i$.

As explained in~\cite{1997PhRvD..55.5895T}, the CMB statistical properties are
fully described by its covariance matrix; thus, in order to prove $\bm{S}_i$
is a properly constrained CMB realization, we need to show that it has the
same covariance matrix with the expected CMB signal, i.e., it is statistically
indistinguishable with a true CMB signal. The mathematical proof of this point
is given as follows: With eq.~(\ref{equ:CR}), the covariance matrix of
$\bm{S}_i$ is
\begin{align}\label{equ:proof pixel domain}
\<\bm{S}_i\bm{S}_i^t\> =& \<(\bm{\widetilde{S}} + \bm{s}_i-\bm{M}\bm{x}_i)\cdot
(\bm{\widetilde{S}} + \bm{s}_i-\bm{M}\bm{x}_i)^t\>
\\ \nonumber
=& \<\bm{\widetilde{S}}\bm{\widetilde{S}}^t\> + \<\bm{s}_i\bm{s}_i^t\> + 
\bm{M}\<\bm{x}_i\bm{x}_i^t\>\bm{M}^t - \bm{M}\<\bm{x}_i\bm{s}_i^t\> - 
\<\bm{s}_i\bm{x}_i^t\>\bm{M}^t
\\ \nonumber
=& \bm{M}\bm{Q}\bm{M}^t + \bm{P} + \bm{M}\bm{Q}\bm{M}^t-\bm{M}\bm{P}-\bm{P}\bm{M}^t
\\ \nonumber
=& \bm{P} + 2\bm{M}\bm{Q}\bm{M}^t - \bm{M}\bm{P} - \bm{P}\bm{M}^t
\\ \nonumber
=& \bm{P} + 2\bm{P}\bm{Q}^{-1}\bm{Q}\bm{Q}^{-1}\bm{P}-2\bm{P}\bm{Q}^{-1}\bm{P}
\\ \nonumber
=& \bm{P}.
\end{align}
Therefore, the constrained realizations $\bm{S}_i$ have identical statistical
properties with the true CMB signal, and is hence a properly constrained
fullsky CMB realization. Moreover, because the CMB angular power spectrum is
completely determined by the pixel domain covariance matrix, we automatically
get the conclusion that, the expected angular power of $\bm{S}_i$ is also
identical to the expected CMB angular power spectrum. Therefore,
eq.~(\ref{equ:CR}) is enough to establish the mathematical basis of MrMVP for
both the pixel and harmonic domains, and the rest of the problem is about the
multi-resolution scheme and details of implementation, which will be described
in the next section.

\section{Implementation and technical details}
\label{sec:implementation}

In this section, we describe some technical details of the implementation of
MrMVP.

\subsection{Fast computation of the pixel domain covariance matrix}
\label{sub:cov fast}

The implementation of eq.~(\ref{equ:CR}) depends on computing the pixel-domain
covariance matrices $\bm{P}$ and $\bm{Q}$ that are consist of the CMB and
non-CMB parts. Using the non-polarized case for illustration, the CMB part of
the covariance matrix is given by the following equation:
\begin{equation}\label{equ:cov}
C_{ij} = \frac{1}{4\pi} \sum_\ell{(2l+1) W_\ell^2 C_\ell P_\ell[\cos{(\theta_{ij})}]},
\end{equation}
where $C_{ij}$ is the element of the CMB covariance matrix, $C_\ell$ is the
CMB APS, $W_\ell$ is the transfer function, $P_\ell(x)$ is the Legendre
polynomial, and $\theta_{ij}$ is the open angle between the $i$-th and $j$-th
pixels. Let's take $N_{side}=512$ as an example: At this resolution, there are
about $3\times10^6$ pixels on the sky, and we usually go up to
$\ell_{max}=1536$. Thus there are about $10^{13}$ elements in the covariance
matrix, and the number of float point operations is at the order of $10^{16}$.
Although eq.~(\ref{equ:cov}) is not very complicated, such a big number of
operations should still be avoided.

In this work, the idea is to first greatly oversample $\theta_{ij}$ from 0 to
$\pi$. In the HEALPix pixelization scheme, at a resolution of $N_{side}$,
$4N_{side}$ points between 0 and $2\pi$ are usually enough to support the
spherical harmonic transforms, but we oversample it to $64N_{side}$ points to
compute the function $C(\theta)$, which is more than enough to guarantee the
precision. Then the full covariance matrix is computed by linear interpolation
of the oversampled $C(\theta)$ function. By this way, the time cost is reduced
from $aN_{pix}^3$ to $bN_{pix}^2$. Note that not only the power index is
reduced from 3 to 2, but also the constant coefficient $b$ is much smaller
than $a$, because it is much easier to compute the interpolation than to
compute the Legendre polynomials.

In case of the polarization, the full pixel domain covariance matrices are
given in eq.~(10) of~\cite{0004-637X-503-1-1} or eq.~(A10-A15) of
~\cite{2001PhRvD..64f3001T}, and they are also in form of $C(\theta)$; thus,
the computation can be accelerated with the same technique.

\subsection{The degree-of-freedom problem}
\label{sub:DOF problem}

According to eq.~(\ref{equ:min-var basic}), we need to invert the covariance
matrix $\bm{Q}$ to get the solution, thus we want $\bm{Q}$ to be non-singular.
For convenience of discussion, $\bm{Q}$ is converted to the harmonic domain:
If the CMB signal is Gaussian and isotropic, then a full covariance matrix of
the CMB in the harmonic domain is diagonal and takes the following look:
\begin{eqnarray}
\bm{C} = 
\begin{pmatrix}
C_0 & 0 & 0 & 0 & 0 & \cdots\\
0 & C_1 & 0 & 0 & 0 & \cdots\\
0 & 0 & C_1 & 0 & 0 & \cdots\\
0 & 0 & 0 & C_1 & 0 & \cdots\\
0 & 0 & 0 & 0 & C_2 & \cdots\\
\cdots & \cdots & \cdots & \cdots & \cdots & \cdots
\end{pmatrix},
\end{eqnarray}
where each $C_\ell$ appears $2\ell+1$ times. Thus for a finite $\ell_{max}$,
the maximum number of non-zero diagonal elements is $(\ell_{max}+1)^2$, which
is equal to the matrix's upper limit of degree-of-freedom (DOF).

However, the size of the covariance matrix is determined not by $\ell_{max}$,
but by the number of map pixels. With the HEALPix pixelization scheme, this
number is equal to $12N_{side}^2$, where $N_{side}=2,4,8,\cdots$ is the
resolution parameter. Therefore, to ensure that $\bm{Q}$ have enough DOF and
is hence non-singular, the following inequality must be satisfied:
\begin{eqnarray}\label{DOF condition}
(\ell_{max}+1)^2 \ge 12N_{side}^2 \Longrightarrow \ell_{max}\ge
\sqrt{12}N_{side}-1\approx 3.5N_{side}.
\end{eqnarray}
However, according to the HEALPix manual, the pixelization scheme by HEALPix
can only support a computation up to $\ell_{max}=3N_{side}-1$. Higher $\ell$s
can still be computed but are no longer reliable. Thus we have a gap between
$3N_{side}$ and $3.5N_{side}$, which can make $\bm{Q}$ singular. Because of
this degree of freedom problem, it is actually impossible to do a pixel domain
minimum variance estimation at the original resolution without prior
information --- it has to be done either with a prior APS or by combining two
neighboring resolutions.

In principle, both ways are allowed: If we use a prior APS, then the analysis
becomes a typical Bayesian estimation and can be further developed into a suit
of minimum variance estimations from map to cosmological parameters. This is
our goal in the following works; however, if we choose to combine two
neighboring resolutions, then the estimation is done in a bind way, which
helps to make the result more robust by providing a crosscheck free from the
prior APS. In this work, we are mainly dealing with the pixel domain
estimation; thus, we shall focus on the latter.

\subsection{The transfer function of downgrading/upgrading}
\label{sub:transfer function}

In order to do the minimum variance estimation in a \emph{blind} way and avoid
the DOF problem, we need to downgrade the input map by one level to
$N_{side}'=0.5N_{side}$, and then upgrade it back to the original resolution
to compute the APS up to $\ell_{max}=2N_{side}=4N_{side}'$. This provides a
precise estimation of the APS up to $4N_{side}'$ for the $N_{side}'$
resolution except for the pixelization effect, because the upgraded map is
morphologically \emph{identical} to the map at $N_{side}'$. However, this also
means, when we perform a blind pixel domain minimum variance estimation at
$N_{side}=512$, the input map resolution should be at least $N_{side}=1024$.

\begin{figure*}[!htb]
  \centering
  \includegraphics[width=0.9\textwidth]{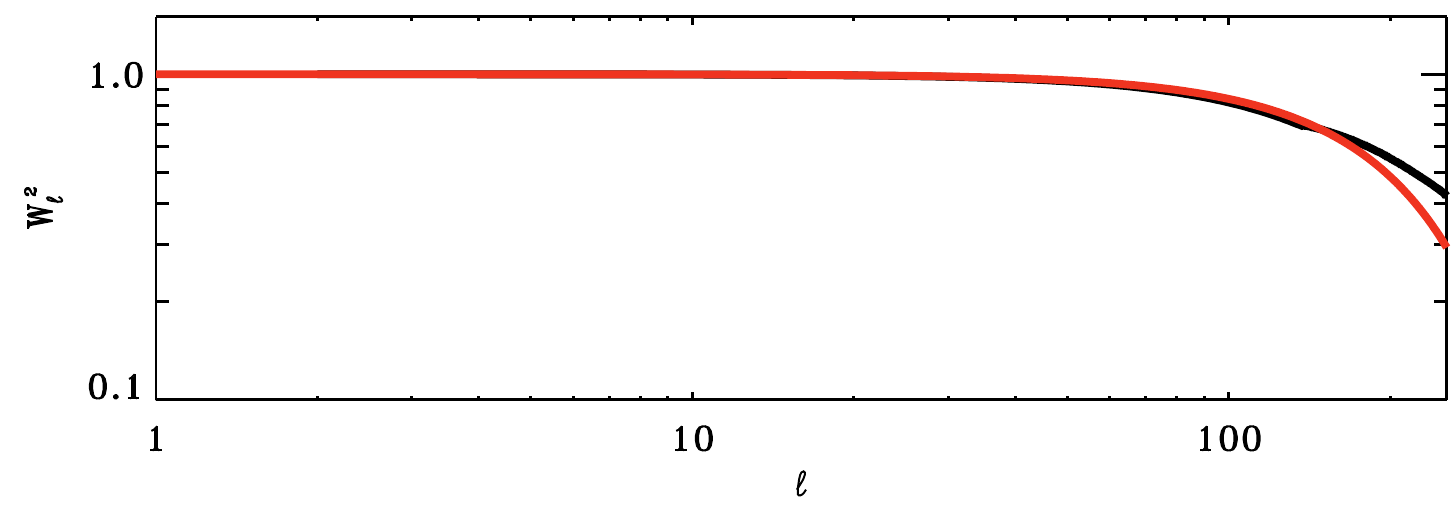}
  \caption{Comparison of the effective transfer function of
  downgrading/upgrading (black) and the one computed directly from the HEALPix
  window functions ignoring the sub-pixel level non-isotropy (red). }
  \label{fig:trans func}
\end{figure*}

The above downgrading/upgrading operation will inevitably change the angular
power spectrum due to pixelization; however, the corresponding effect cannot
be computed by the HEALPix pixel window functions, because a
downgraded-and-upgraded map has identical values for the sub-pixels of
$N_{side}'$ but different values for other pixels, which is non-isotropic at
the sub-pixel level. Therefore, the effective transfer function should be
computed by simulation. Fortunately, this kind of transfer function does not
change significantly with the expected APS and does not need to be updated
frequently. In figure~\ref{fig:trans func}, we compare the transfer functions
given by simulation and the one computed directly from the HEALPix pixel
window function without considering the sub-pixel non-isotropy, with
$N_{side}'=64$. One can see that they are almost identical at
$\ell<2N_{side}'$, but are slightly different at higher $\ell$s.

\subsection{The multi-resolution scheme}
\label{sub:multi-reso}

\begin{figure*}[!htb]
  \centering
  \includegraphics[width=0.48\textwidth]{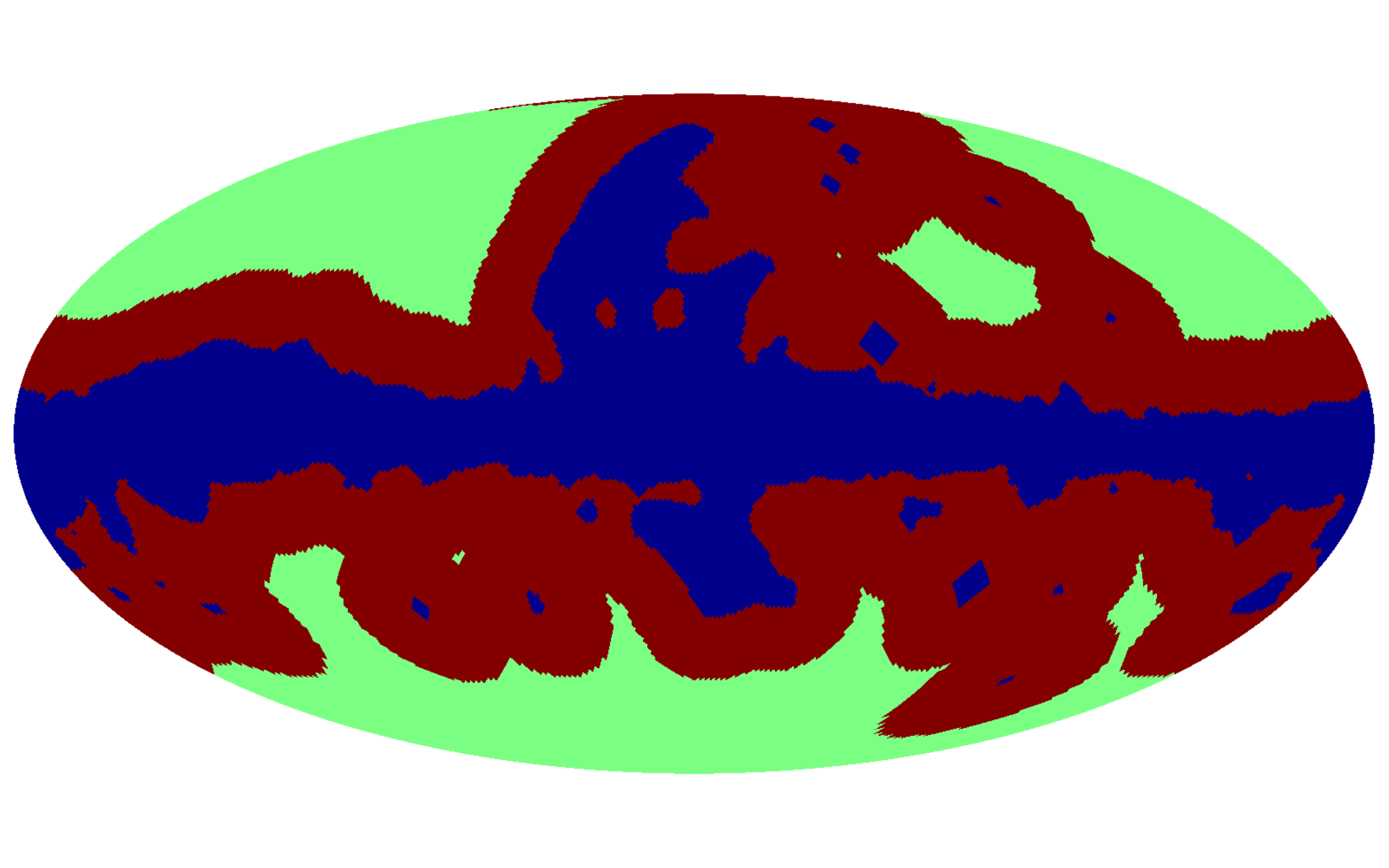}
  \includegraphics[width=0.48\textwidth]{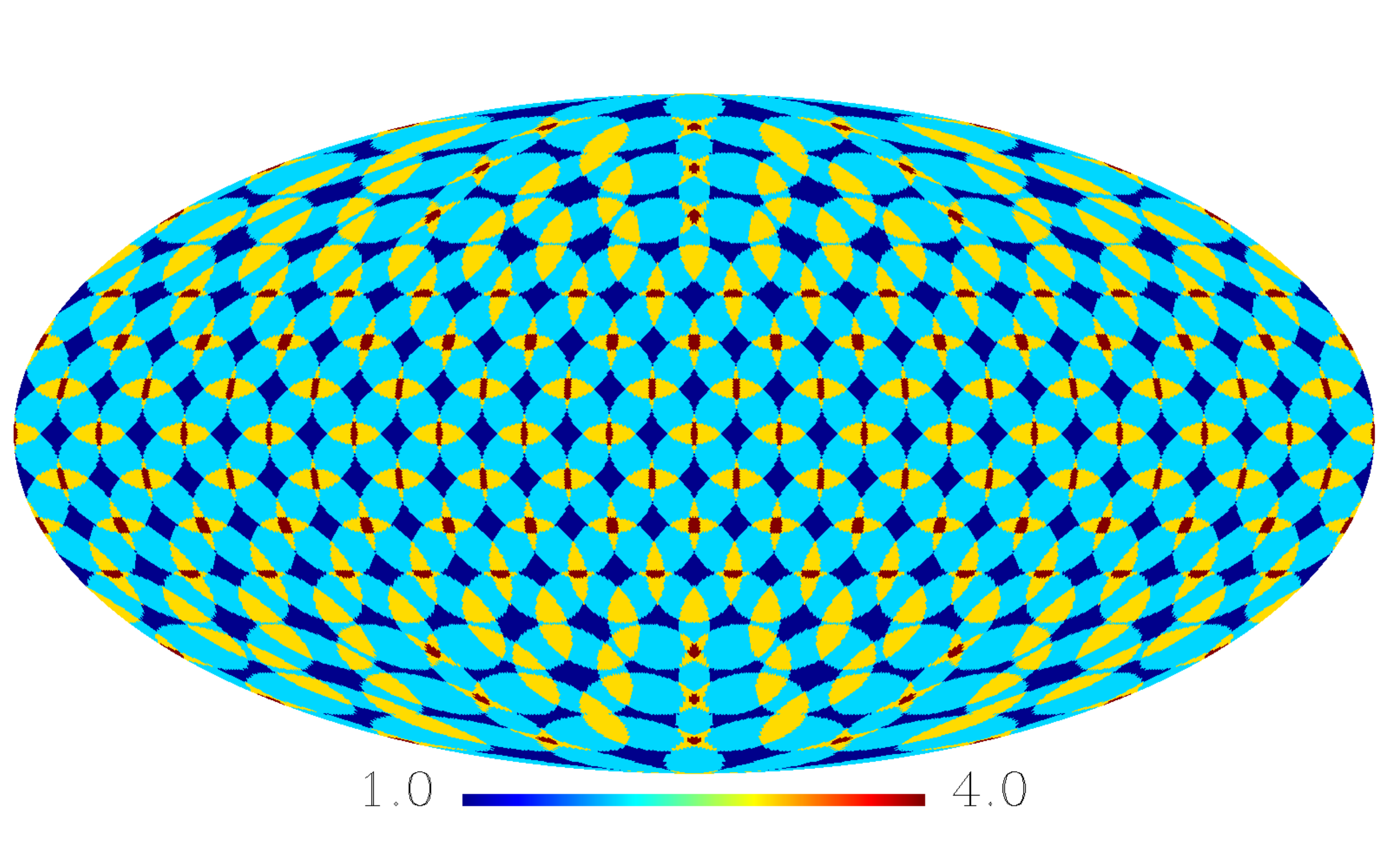}
  \caption{Illustration of the multi-resolution scheme for the missing region
  (left) and for the available region (right). In the left panel, the deep
  blue region is an example of the missing region (WMAP polarization mask),
  and the red area is the region used for reconstruction at working
  $N_{side}=64$. In the right panel are the overlapping mosaic discs at
  working $N_{side}=128$. When the resolution increases, the area in use
  decreases but the number of mosaic discs increases. In any case, each map
  pixel is covered at least once.}
  \label{fig:multi-reso scheme}
\end{figure*}

A full implementation of eq.~(\ref{equ:CR}) at a high resolution requires too
much time and memory, thus we use a multi-resolution scheme to make the
computation feasible. Basically, this means to do a full implementation of
eq.~(\ref{equ:CR}) at $N_{side}=16$, and then go to a higher resolution and
use less pixels for estimation. The schemes are different for the missing and
available regions: For reconstructing the missing region at $N_{side}>16$,
only the available pixels within a certain angular distance from the edge of
the mask are used, and for each higher resolution, this angular distance is
reduced by 50\%. For estimating the available sky region at $N_{side}>16$, the
computations are done in units of mosaic disks centered at the sky pixels of
resolution $N^{md}_{side}=N_{side}/32$, and the radius are
$1536\degree/N_{side}$. These values ensure that the mosaic disks will cover
all sky pixels, but will not be too big to handle. In
figure~\ref{fig:multi-reso scheme}, the sky area used to estimate the
missing/available regions are illustrated, respectively.

In the above schemes, for the case of estimating the missing region, the
number of available pixels in use scales linearly with $N_{side}$, whereas the
number of pixels to be estimated follows $N_{side}^2$; thus, the time cost
follows $N_{side}^3$ for both the matrix inversion and matrix multiplication.
As for the case of estimating the available region, the number of available
pixels in use is nearly constant, but the number of mosaic discs follows
$N_{side}^2$. Therefore, the overall time cost roughly follows $N_{side}^3$.

\subsection{Lossless synthesis of multiple resolutions}
\label{sub:synthesis}

By working with eq.~(\ref{equ:CR}) from $N_{side}=16$ to $N_{side}'$, we get
fullsky estimates of the CMB signal at a series of resolutions. These results
should be synthesized to make a final output. The synthesis can be done either
in the pixel domain or in the harmonic domain. Analysis and practical tests
show that the pixel domain synthesis is not only lossless but also faster,
more stable and more accurate; thus it is the default choice.

With the results at a series of resolutions starting from $N_{side}=16$, the
pixel domain synthesis is done as follows:
\begin{enumerate}
\item\label{itm:syn1} First upgrade the result at $N_{side}=16$ to
$N_{side}=32$, which represents larger scale structures.
\item\label{itm:syn2} Then downgrade the result of $N_{side}=32$ to
$N_{side}=16$ and upgrade back to $N_{side}=32$.
\item\label{itm:syn3} Subtract the product of step~\ref{itm:syn2} from the
result at $N_{side}=32$ to get the difference, which represents smaller scale
structures.
\item\label{itm:syn4} Add the product of step~\ref{itm:syn3} to the product of
step~\ref{itm:syn1} to make the synthesized result of $N_{side}=32$.
\item Start again from step~\ref{itm:syn1} with the product of
step~\ref{itm:syn4} to synthesize the next (higher) resolution, until reaching
$N_{side}'$.
\end{enumerate}
Practically, the above procedures can also be done in a descending order from
$N_{side}'$ to $N_{side}=16$ to save the memory, and in this way, the original
result at a higher resolution can be deleted after each round to free some
memory.

It is easy to prove that the above synthesis procedures are lossless, i.e.,
when we do it with a series of multi-resolution maps downgraded from the same
top-resolution map, the original map is restored by 100\%. This also means the
above synthesis procedures are safe for both temperature and polarizations,
and will not cause a further E-to-B leakage \emph{by itself}.

\subsection{Matrix based fine tuning of the multi-resolution synthesis}
\label{sub:fine tune of synthesis}

The multi-resolution synthesis described in section~\ref{sub:synthesis} has
the following matrix form: In the HEALPix pixelization scheme, each pixel at a
lower resolution contains exactly four smaller pixels of the next (higher)
resolution. Let the CMB signals be $S_0$ at this bigger pixel (lower
resolution) and $S_{1-4}$ at the four smaller pixels (higher resolution), then
the synthesis scheme in section~\ref{sub:synthesis} between two neighboring
resolutions is described by the following matrix equation:
\begin{align}
\begin{pmatrix}
S_1 \\ S_2 \\ S_3\\ S_4
\end{pmatrix}_{syn} = \frac{1}{4}\left[
\begin{pmatrix}
+3 & -1 & -1 & -1 \\
-1 & +3 & -1 & -1 \\
-1 & -1 & +3 & -1 \\
-1 & -1 & -1 & +3 
\end{pmatrix}\cdot
\begin{pmatrix}
S_1 \\ S_2 \\ S_3\\ S_4
\end{pmatrix} +
\begin{pmatrix}
1 & 1 & 1 & 1 \\
1 & 1 & 1 & 1 \\
1 & 1 & 1 & 1 \\
1 & 1 & 1 & 1 
\end{pmatrix}\cdot
\begin{pmatrix}
S_0 \\ S_0 \\ S_0\\ S_0
\end{pmatrix}\right],
\end{align}
which can be simplified to
\begin{align}
\begin{pmatrix}
S_1 \\ S_2 \\ S_3\\ S_4
\end{pmatrix}_{syn} = \frac{1}{4}
\begin{pmatrix}
+4 & +3 & -1 & -1 & -1\\
+4 & -1 & +3 & -1 & -1\\
+4 & -1 & -1 & +3 & -1\\
+4 & -1 & -1 & -1 & +3
\end{pmatrix}\cdot
\begin{pmatrix}
S_0 \\ S_1 \\ S_2 \\ S_3 \\ S_4
\end{pmatrix} = 
\bm{M}_{syn}\cdot
\begin{pmatrix}
S_0 \\ S_1 \\ S_2 \\ S_3 \\ S_4
\end{pmatrix}
\end{align}
If the self-consistency condition $S_0=(S_1+S_2+S_3+S_4)/4$ is satisfied, then
$\bm{S}_{syn}=\bm{S}$, which means the synthesis scheme is lossless by itself.

However, the lossless scheme is not unique, and it is possible to design other
lossless synthesis schemes by fine tuning this $4\times5$ synthesis matrix
$\bm{M}_{syn}$ to achieve a better effect. Unfortunately, this job is very
time consuming; thus we leave it for the next work of this series.

\subsection{Pre-processing of tiny holes in the mask}
\label{sub:mask}

The multi-resolution scheme in section~\ref{sub:multi-reso} does not work very
well when the mask contains too many tiny holes, because each tiny hole will
force to use some area around it for reconstruction, which is not really
necessary. Therefore, if the mask contains too many tiny holes, like the ones
for the point sources, then it is recommended to erase those tiny holes by the
diffusive in-paint method~\citep{2014A&A...571A..24P, 2016A&A...594A..17P}.
Although the diffusive in-paint method does not satisfy the minimum variance
condition, it is especially safe and convenient for tiny holes in a mask, and
the corresponding precision loss is usually negligible compared to the help it
brings to us.

\subsection{About the input CMB APS}
\label{sub:input APS}

The implementation of eq.~(\ref{equ:CR}) requires to use the CMB covariance
matrix, which is determined by the input CMB APS; however, if we start from
the pixel domain map directly, there is no CMB APS in advance. Therefore, we
should either use a prior CMB APS, which turns the work into a typical
Bayesian analysis; or we should assume an initial CMB APS, and improve it by
the expectation of APS within $\{\bm{S}_i\}$ to get an iterative solution,
which leads to a blind solution that does not depend on the prior APS. Both
ways are allowed, and for the latter, the number of rounds needed to converge
is roughly a few times of $1/f_{sky}$, which means when the available region
is very small, the convergence can be slow. The reason for that is: when the
available region is very small, it cannot provide enough correction to the
fullsky APS in one round. However, in this case, the reconstruction of the
missing region becomes much less important, so we can consider stopping at a
lower resolution for the missing region reconstruction, and focus on the
estimation of the available region, which can make the computation much
faster. Another possible idea is to amplify the modification to the resulting
APS by a certain factor to accelerate the convergence for a small $f_{sky}$.
Both ways will be tested in the following works.

\section{Test results}
\label{sec:results}

We start testing MrMVP in the temperature case without noise at
$N_{side}=128$, with an input $\Lambda$CDM power spectrum from the most recent
Planck collaboration release \cite{2018arXiv180706209P}, and the mask is
chosen to be the WMAP KP2 mask without the point sources. With a 4-core
laptop, we generate 1,000 realizations in the constrained ensemble
$\{\bm{S}_i\}$, and the total time cost is illustrated in figure~\ref{fig:time
cost} as a function of the resolution, which follows $O(N_{side}^3)$ nicely.
According to the figure, the average time cost to generate one realization at
$N_{side}=128$ is about 0.1 second, which is a significantly improved speed
compared to previous works.
\begin{figure*}[!htb]
  \centering
  \includegraphics[width=0.96\textwidth]{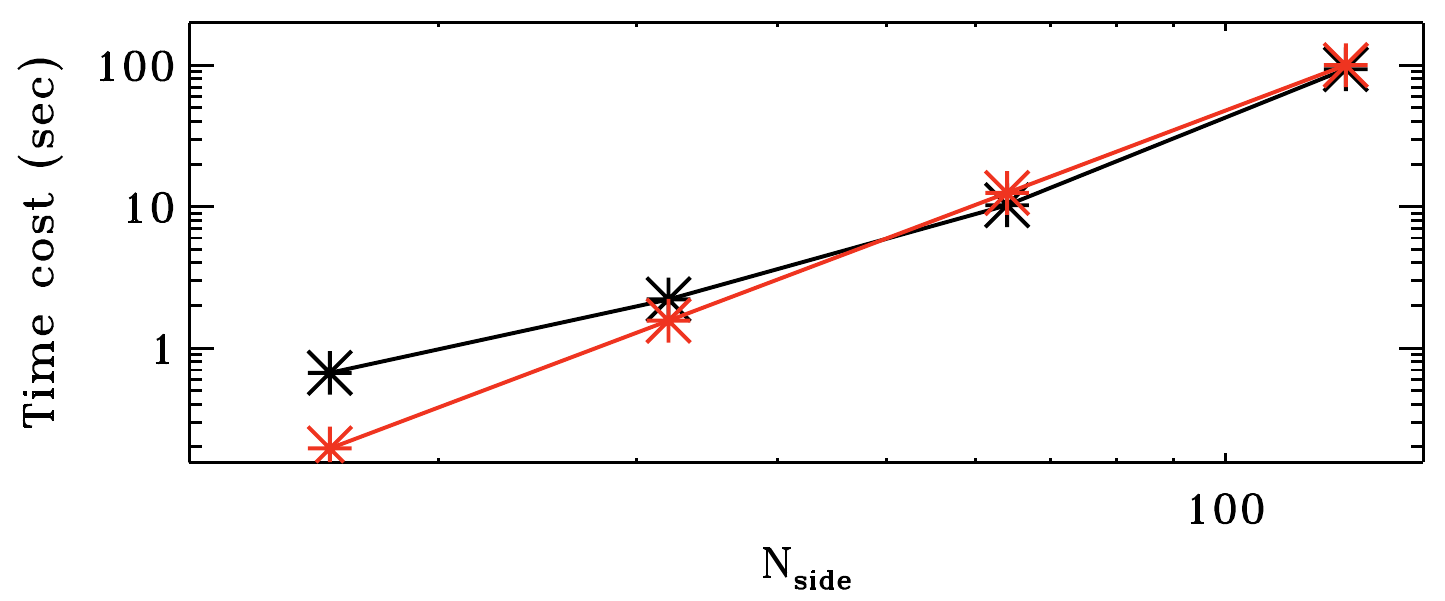}
  \caption{Illustration of the total time cost of generating 1000 realizations
  in $\{\bm{S}_i\}$ (black) from $N_{side}=16$ to $N_{side}=128$. The red line
  is for the $O(N_{side}^3)$ expectation. At low resolutions, the total time
  cost is higher than $O(N_{side}^3)$ due to the auxiliary operations like
  disk I/O, consistency test, mask and region processing, etc.}
  \label{fig:time cost}
\end{figure*}

\begin{figure*}[!htb]
  \centering
  \includegraphics[width=0.48\textwidth]{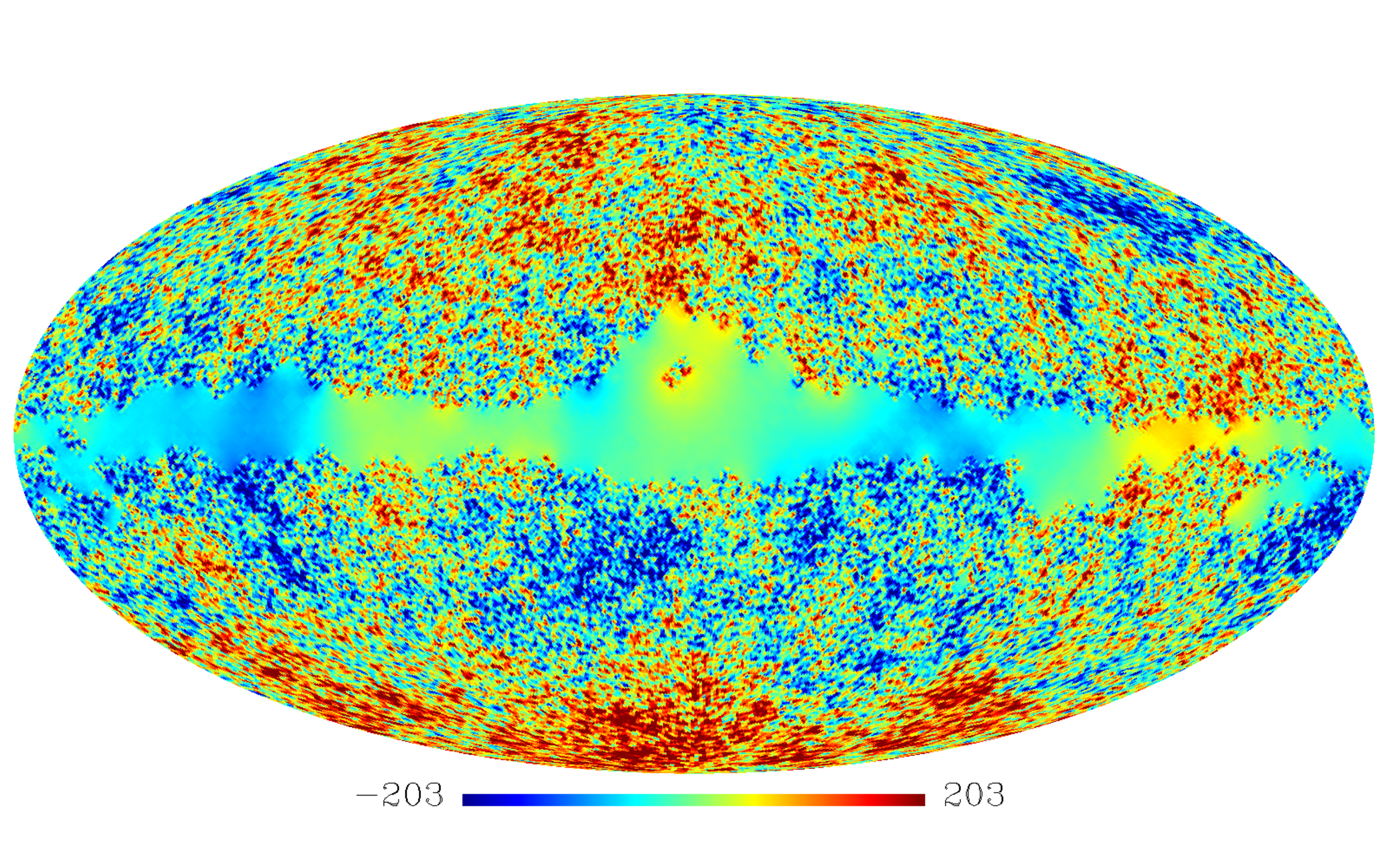}
  \includegraphics[width=0.48\textwidth]{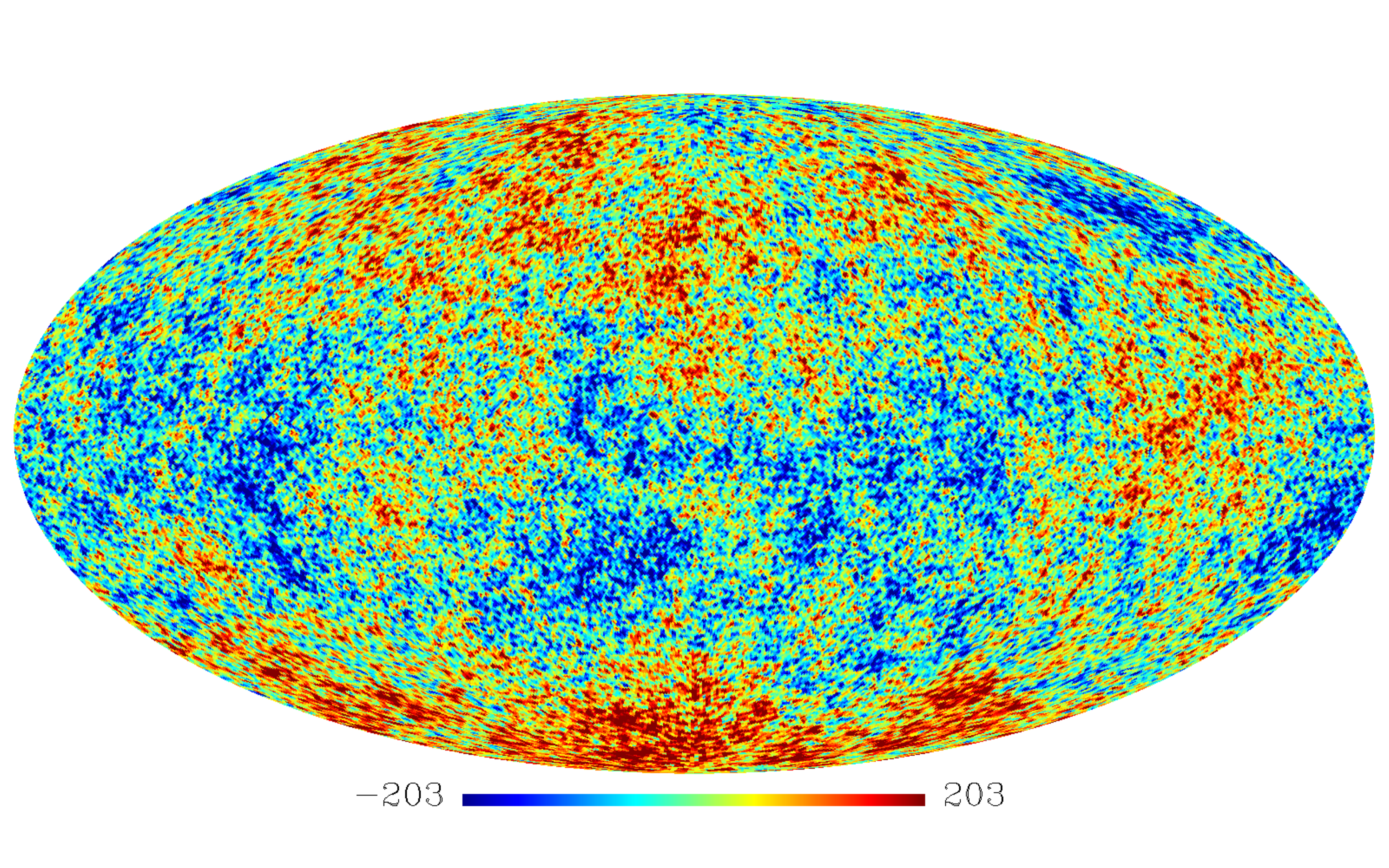}

  \includegraphics[width=0.48\textwidth]{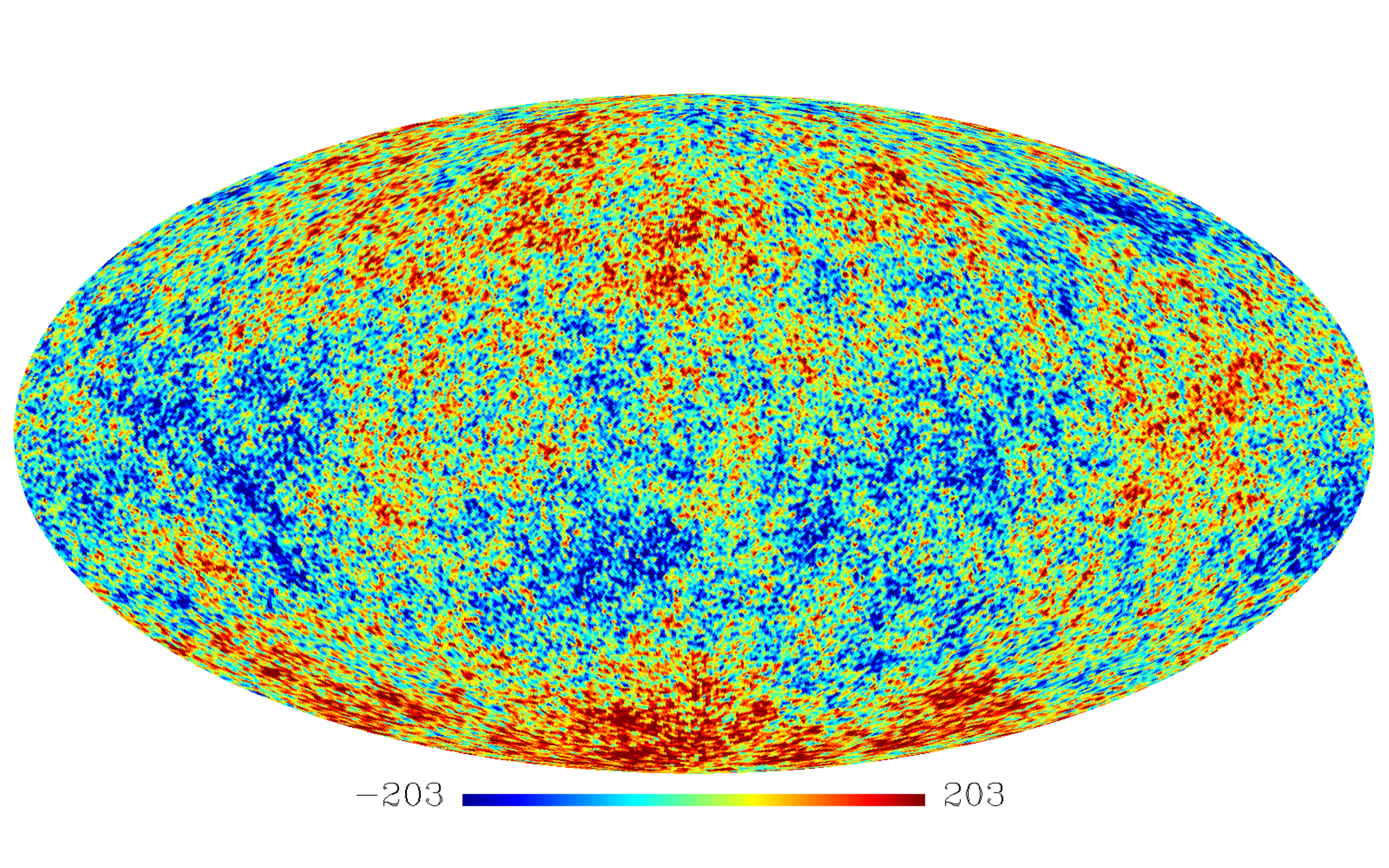}
  \includegraphics[width=0.48\textwidth]{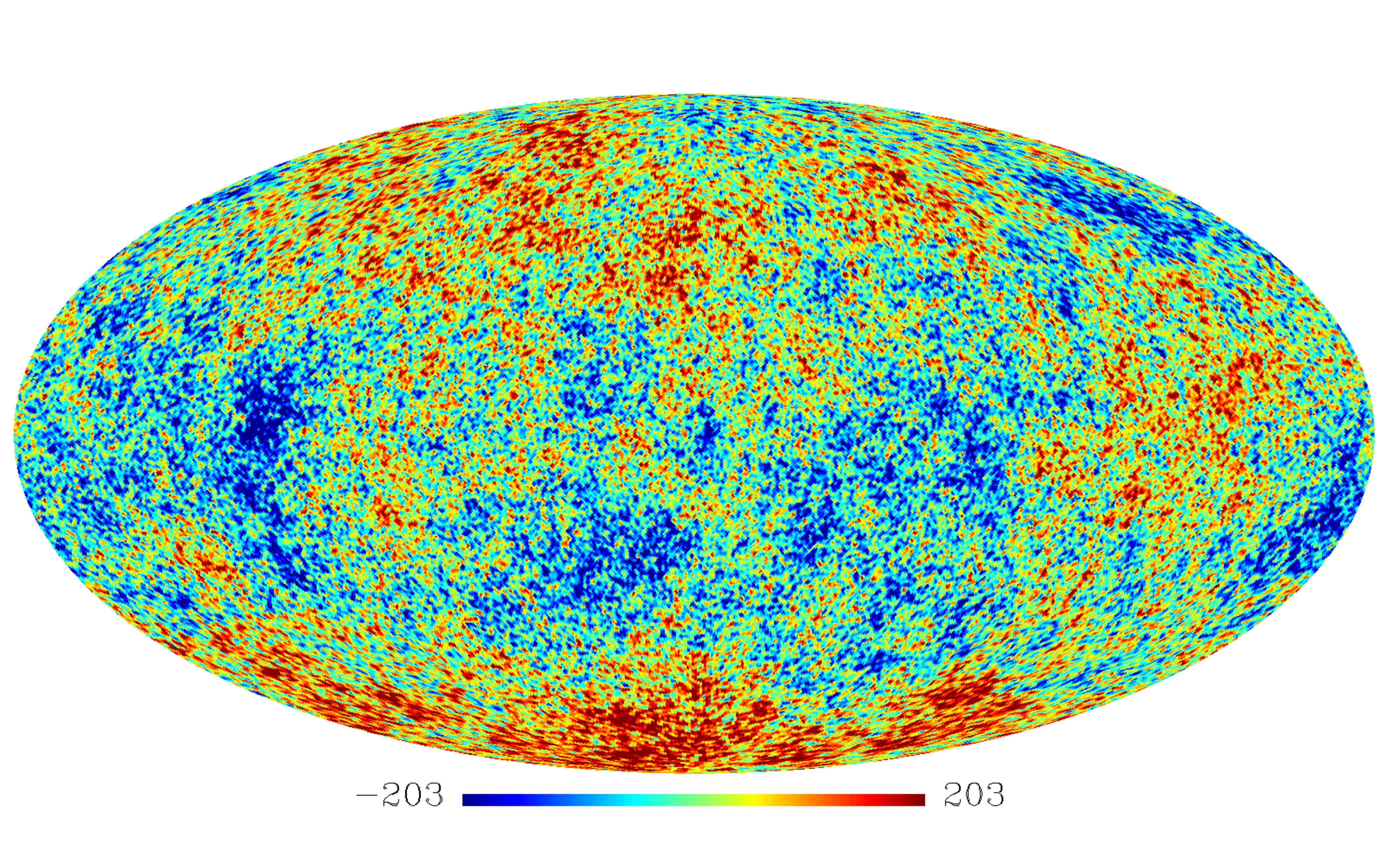}
  \caption{Comparison of the pixel domain expectation $\widetilde{\bm{S}}$
  (upper-left), the input simulated CMB map (upper right), and two more
  constrained realizations $\bm{S}_i$ derived by pixel domain operations
  (lower panels).}
  \label{fig:map comparison}
\end{figure*}

Then we compare the pixel domain expectation $\widetilde{\bm{S}}$, the input
pixel domain CMB map and two constrained realizations $\bm{S}_i$ in
figure~\ref{fig:map comparison}. Note that $\widetilde{\bm{S}}$ is the minimum
variance solution in the pixel domain, but its APS is not a minimum variance
estimate of the CMB APS. The latter should be derived from the expectation of
APS within $\{\bm{S}_i\}$.

\begin{figure*}[!htb]
  \centering
  \includegraphics[width=0.96\textwidth]{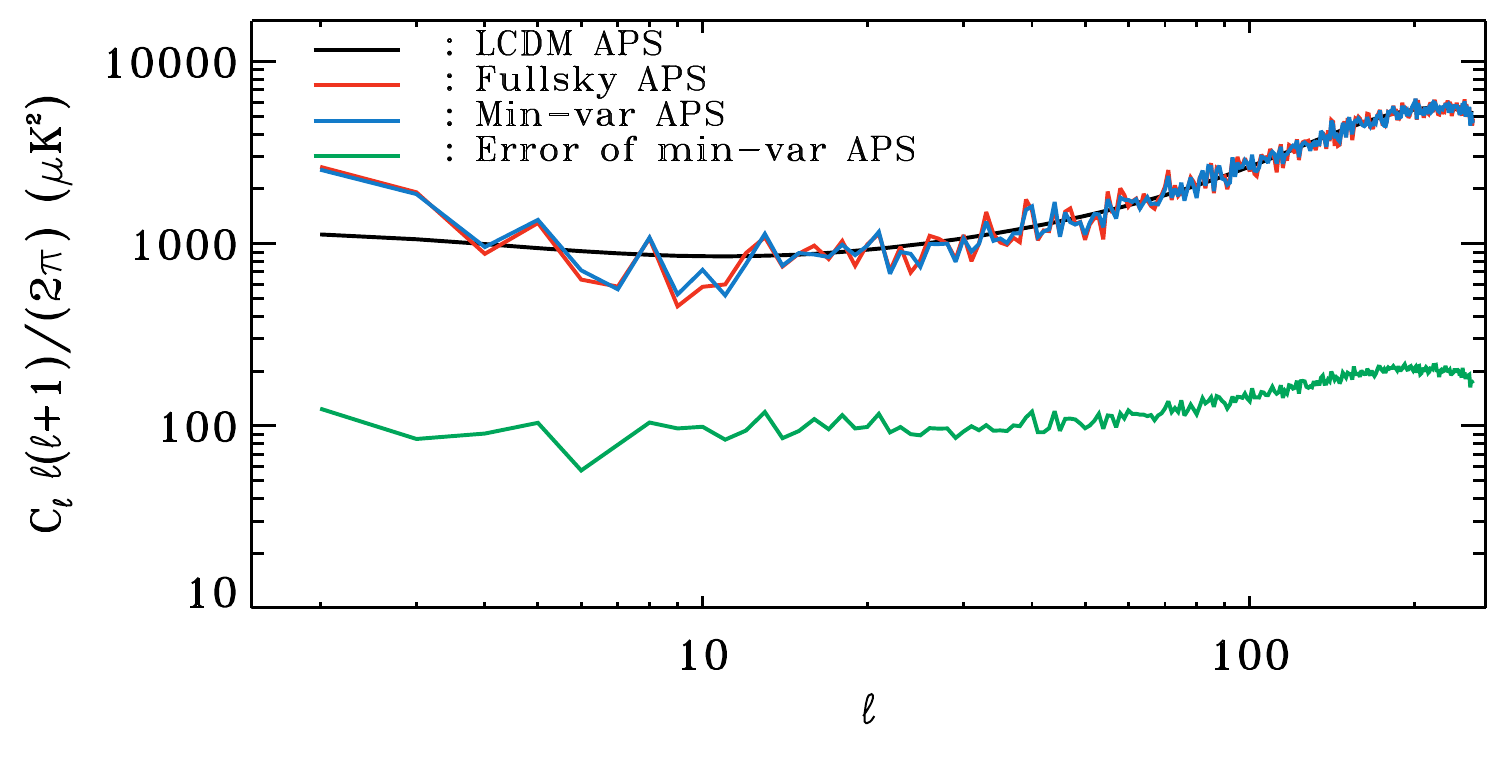}
  \caption{Comparison of the APS of the: input $\Lambda$CDM (black), fullsky
  CMB map (red), minimum variance estimate (blue); and the RMS error of the
  minimum variance estimate (green).}
  \label{fig:dl comparison}
\end{figure*}

In figure~\ref{fig:dl comparison}, we compare the input $\Lambda$CDM APS, the
APS of the fullsky CMB map, the minimum variance estimate of the APS, and the
RMS error of the minimum variance estimate. We can see that the minimum
variance estimate of the APS is very well consistent with the fullsky CMB
power spectrum.

Note that, because the entire ensemble $\{\bm{S}_i\}$ are constrained, the
corresponding minimum variance estimate and the error amplitudes are also
constrained. Therefore, the APS error bars derived from $\{\bm{S}_i\}$ are
more informative than a general estimation of the error bars without
considering the constraint. For example, the error bar at $\ell=2$ will follow
the amplitude of the true quadrupole power; thus, if the true quadrupole power
happens to be quite high (like the one in figure~\ref{fig:dl comparison}),
then its error bar will increase accordingly, which faithfully reflects the
actual constraint from the available dataset.

\begin{figure*}[!htb]
  \centering
  \includegraphics[width=0.24\textwidth]{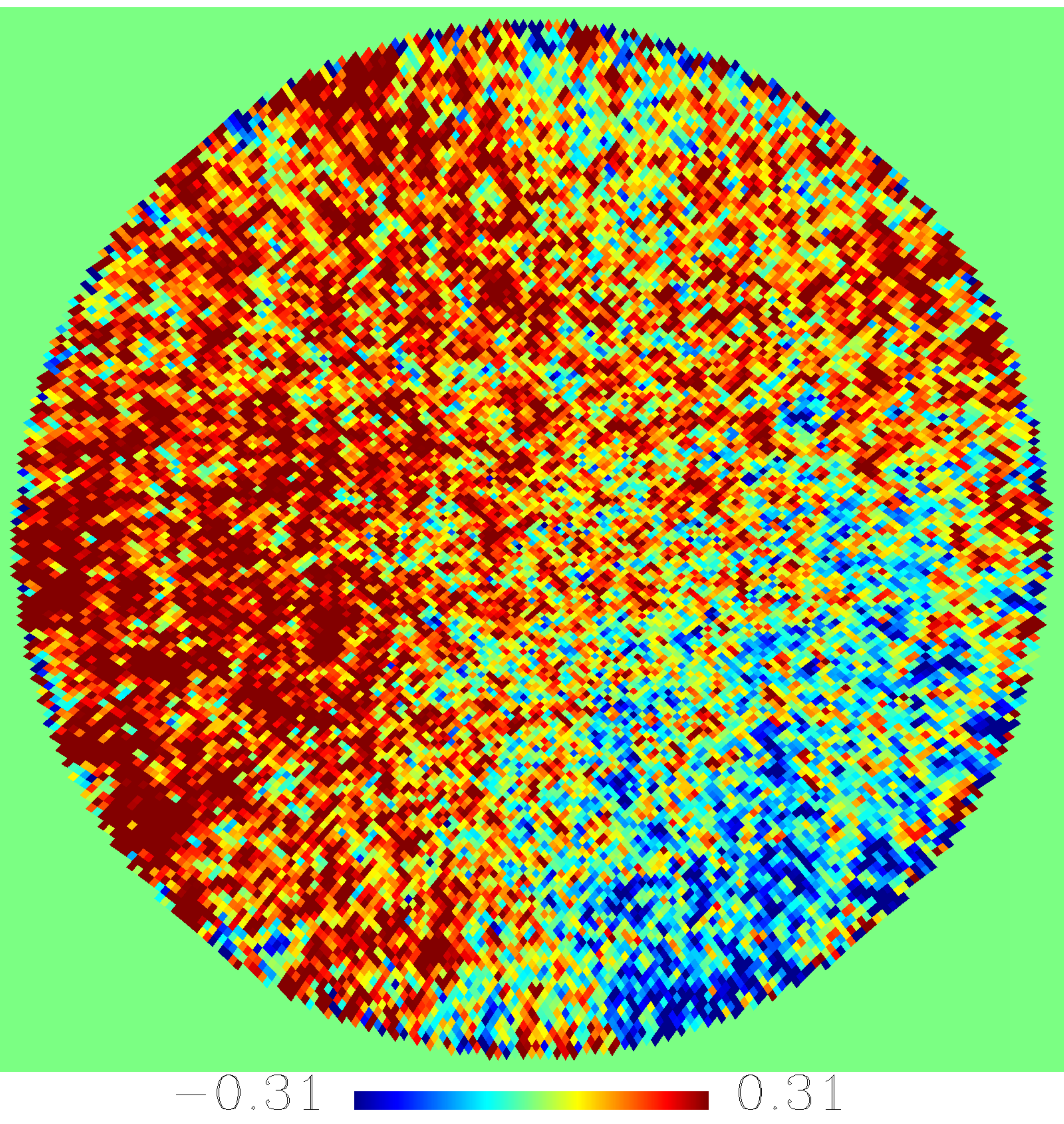}
  \includegraphics[width=0.24\textwidth]{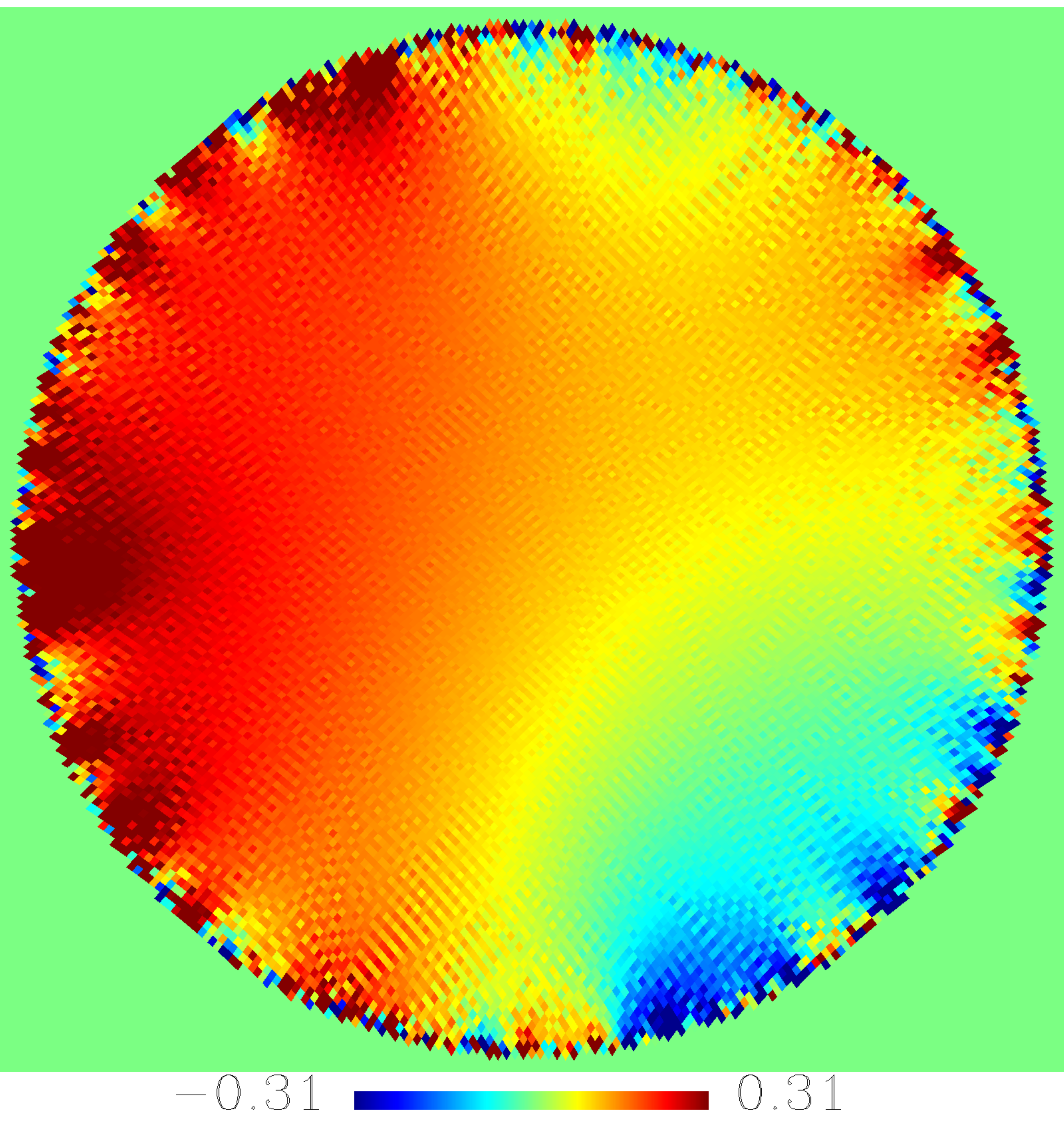}
  \includegraphics[width=0.24\textwidth]{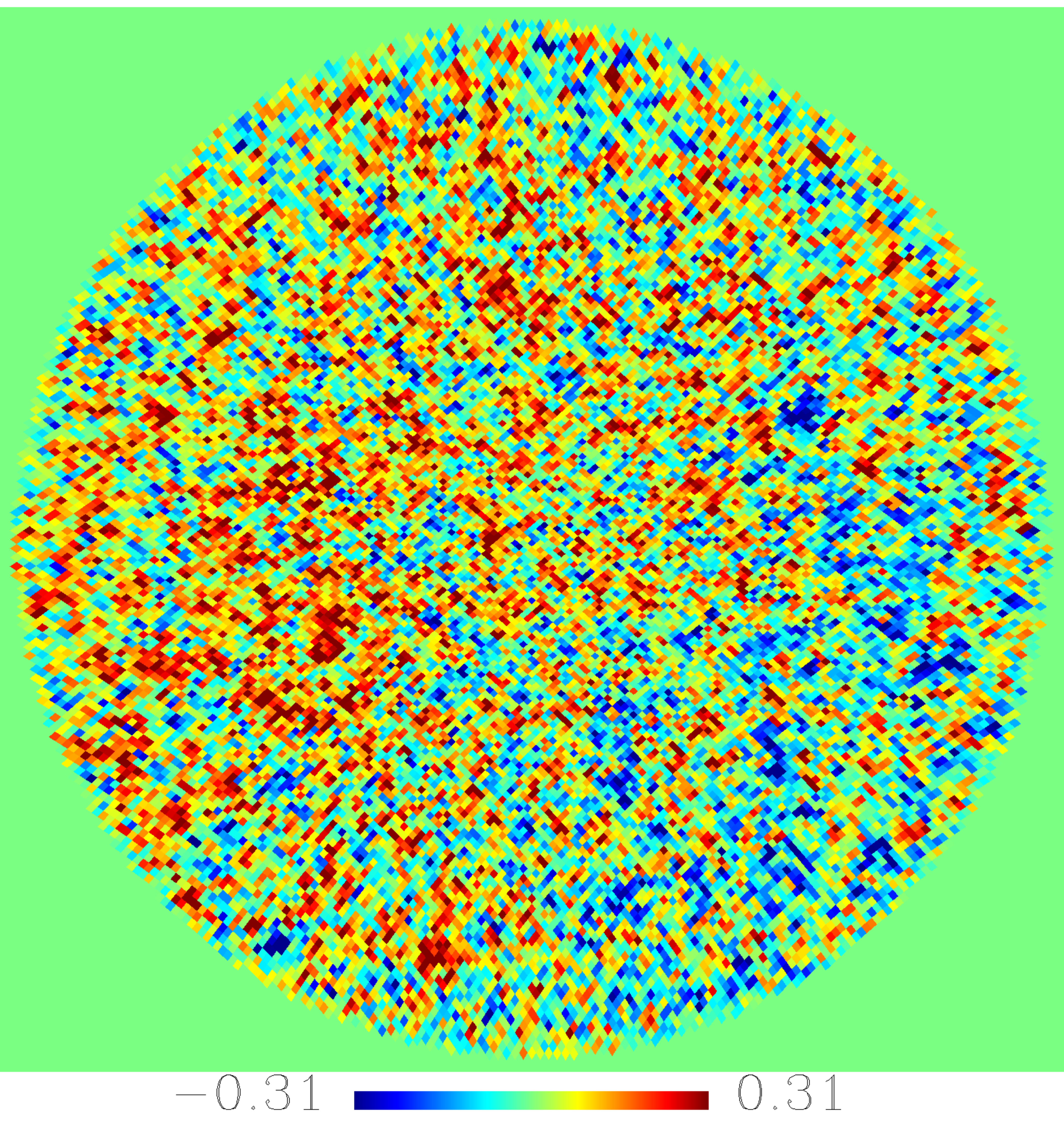}
  \includegraphics[width=0.24\textwidth]{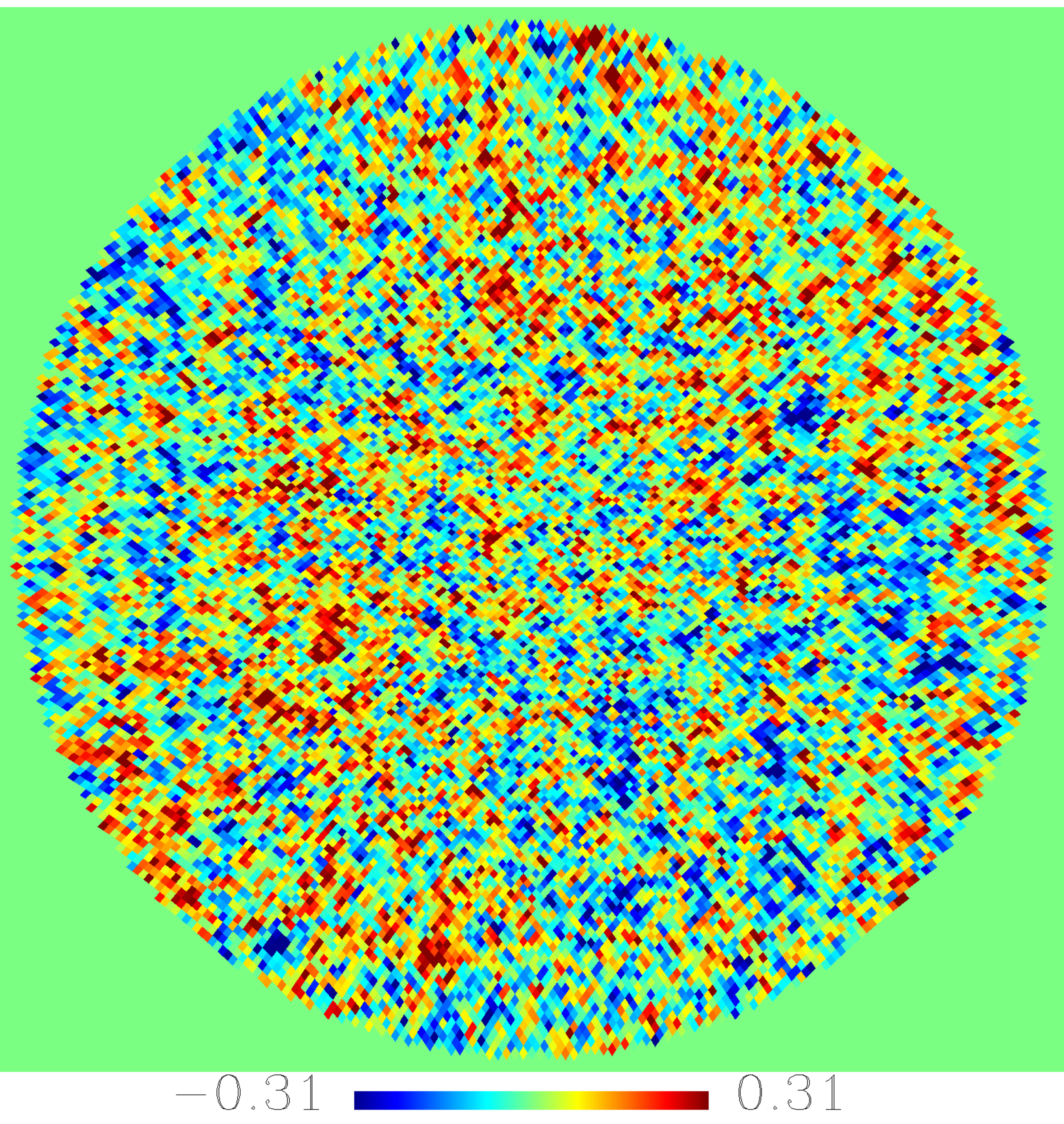}
  \caption{Illustration of the effectiveness of the minimum variance estimate
  of the available region. Panel 1 (from left to right) is an available region
  that is contaminated by the EB-leakage residuals after correction, and the
  residual component itself is shown in panel 2. After the minimum variance
  estimation, we obtain a clean region (panel 3), which is significantly
  improved compared to the contaminated one, and is better consistent with the
  true input (panel 4). The unit is $\mu$K.}
  \label{fig:available region}
\end{figure*}

For the available region, the effect of MrMVP is to alleviate the unwanted
residuals. The effectiveness of alleviation is shown in
figure~\ref{fig:available region}, where the input signal is a simulated
B-mode map with $r=0.05$, the available region is a disk in the center of the
sky, and the contaminant is the residual of EB-leakage correction following
the blind template correction method introduced in~\cite{2018arXiv181104691L,
2019arXiv190400451L, Liu_2019_EB_general}. Such a residual is significantly
correlated between different map pixels and is significantly non-isotropic. To
make the residuals more visible and to test the method in a more difficult
situation, the simulated residuals are amplified by factor 2, so they are
bigger than what we may actually encounter. As we see from
figure~\ref{fig:available region}, the minimum variance estimate of the
available region can efficiently alleviate the residual contamination, as far
as its covariance matrix can be estimated.

In summary, figures~\ref{fig:map comparison}--\ref{fig:dl comparison} confirm
the correctness of the MrMVP implementation, figure~\ref{fig:time cost} shows
that it is very fast and scaled only as $O(N_{side}^3)$, and
figure~\ref{fig:available region} shows that it can efficiently alleviate the
residual errors with known covariance matrices, even if they are correlated
and non-isotropic, like the residual of EB-leakage correction.

\section{Conclusion and discussion}
\label{sec:conclusion}

In this work, we have designed a pixel domain multi-resolution minimum
variance painting method (MrMVP). Its time cost follows $O(N_{side}^3)$, which
is much better than the $O(N_{side}^6)$ time cost of a traditional minimum
variance method. The method helps to generate full sky CMB realizations that
are properly constrained by the observed sky map, and a mathematical proof is
given to confirm that these constrained realizations have identical
statistical properties with the expected CMB signal, which is also consistent
with the test results. When the method is applied to the missing region, it
helps to alleviate the incomplete sky effect, and when it is applied to the
available region, it can efficiently alleviate the unwanted pixel domain
residuals with known covariance matrices, even if they are significantly
correlated and non-isotropic, like the residual of EB-leakage. Both sides are
useful in the data analysis of CMB experiments.

However, the method still need to be further developed and improved:
additional routines needs to be developed to make it suitable for a complete
Bayesian analysis from the pixel domain map to the final cosmological
parameters; and the method itself needs to be fine tuned, especially for the
synthesis matrix, to further improve its performance. Meanwhile, a GPU
implementation of the method is also under consideration, because it can
significantly reduce the time cost of inverting the covariance matrices. For
small $f_{sky}$, several ideas of improving the convergence speed are proposed
and will be tested in the following works.

\Ack{

This research has made use of the \textsc{HEALPix}~\citep{2005ApJ...622..759G}
package and data product from the Planck~\citep{Planckdata:online}
collaboration. Hao Liu is also supported by the Youth Innovation Promotion
Association, CAS.

}



\providecommand{\href}[2]{#2}\begingroup\raggedright\endgroup

\end{document}